\documentclass{article}

\usepackage{PRIMEarxiv}

\usepackage[utf8]{inputenc} 
\usepackage[T1]{fontenc}    
\usepackage{hyperref}       
\usepackage{url}            
\usepackage{booktabs}       
\usepackage{amsfonts}       
\usepackage{nicefrac}       
\usepackage{microtype}      
\usepackage{lipsum}
\usepackage{fancyhdr}       
\usepackage{graphicx}       

\usepackage[export]{adjustbox}

\usepackage[english]{babel}

\usepackage{doi}

\usepackage{comment}  

\usepackage{amsmath} 

\usepackage{amssymb} 

\usepackage{amsthm}

\usepackage[ruled,linesnumbered]{algorithm2e}

\usepackage{float}

\usepackage{subcaption}

\usepackage{silence}

\usepackage{dirtytalk}

\newtheorem{theorem}{Theorem}[section]
\newtheorem{lemma}[theorem]{Lemma}

\graphicspath{{media/}}     

\title{A Committee Based Optimal Asynchronous Byzantine Agreement Protocol W.P. 1}

\author{ {Nasit S Sony} \\
	University of California, Merced\\
	CA 95340, USA \\
	\texttt{nsony@ucmerced.edu} \\
	\And
	{Xianzhong Ding} \\
	Lawrence Berkeley National Laboratory\\
	CA 94720, USA \\
	\texttt{dingxianzhong@lbl.gov} \\
        \And
	{Mukesh Singhal} \\
	University of California, Merced\\
	CA 95340, USA \\
	\texttt{msinghal@ucmerced.edu} \\
}

\begin{document}
\maketitle

\begin{abstract}
Multi-valued Byzantine agreement (MVBA) protocols are essential for atomic broadcast and fault-tolerant state machine replication in asynchronous networks. Despite advances, challenges persist in optimizing these protocols for communication and computation efficiency. This paper presents a committee-based MVBA protocol (cMVBA), a novel approach that achieves agreement without extra communication rounds by analyzing message patterns in asynchronous networks with probability $1$. Unlike traditional MVBA, which requires all \( n \) parties (where \( n = 3f+1 \), with \( f \) as the maximum number of faulty parties), cMVBA leverages a committee-based selection where a subset of \( f+1 \) parties—ensuring at least one honest party—can efficiently achieve agreement. Integrating the asynchronous binary Byzantine agreement protocol, cMVBA uses verifiable proofs from these parties to finalize the agreement. The protocol is resilient to up to \( \lfloor n/3 \rfloor \) Byzantine failures, with an expected runtime of \( O(1) \), message complexity of \( O(n^2) \), and communication complexity of \( O((l + \lambda)n^2) \), where \( l \) is the input bit length and \( \lambda \) the security parameter.
\end{abstract}


\section{Introduction}
\label{sec:intro}

Byzantine Agreement (BA) protocols are essential for agreement in decentralized infrastructures, particularly in blockchain and other distributed applications. These protocols ensure agreement even with malicious actors, a necessity highlighted by the success of applications like Bitcoin \cite{BITCOIN01} and similar decentralized systems \cite{ BLOCKCHAIN01}. However, existing asynchronous BA protocols face high communication costs and require time parameters to guarantee liveness, which limits their practical efficiency. The FLP impossibility result \cite{CONS03} further demonstrates that deterministic BA protocols cannot achieve guaranteed agreement in asynchronous networks, creating a need for efficient alternatives \cite{BYZ17}.

Traditional Multi-Valued Byzantine Agreement (MVBA) protocols like Cachin's \cite{CACHIN01} achieve agreement over large inputs but are constrained by high communication complexity, typically \(O(n^3)\), making them impractical for large-scale systems. Approaches like VABA \cite{BYZ17} and Dumbo-MVBA \cite{BYZ20} improve efficiency with view-based methods and erasure coding, but at the expense of additional communication rounds. Signature-free solutions, such as Mostefaoui et al.’s protocol \cite{SIG01}, further reduce overhead by relaxing validity conditions, yet achieving strong validity in asynchronous settings remains challenging.

We propose a Multi-Valued Byzantine Agreement (cMVBA) protocol to address these limitations. Our approach introduces a committee-based structure that dynamically selects a subset of \(f+1\) parties, ensuring at least one honest party in each instance. This design reduces message complexity by integrating the Asynchronous Binary Byzantine Agreement (ABBA) protocol, which reaches agreement with probability 1 without extra communication rounds. The cMVBA protocol achieves optimal performance with runtime \(O(1)\), message complexity \(O(n^2)\), and communication complexity \(O((l + \lambda)n^2)\), where \(l\) is the input bit length and \(\lambda\) the security parameter.

\textbf{Contributions}:
\begin{itemize}
    \item Introduce a committee selection protocol with at least one honest party, enhancing resilience against Byzantine failures.
    \item Propose cMVBA, reducing communication complexity from \(O(n^3)\) to \(O((l + \lambda)n^2)\) by selectively broadcasting proposals.
    \item Integrate ABBA to achieve reliable agreement on a proposal by a committee member.
    \item Provide theoretical analysis showing the effectiveness and the correctness of the protocol.
\end{itemize}

\begin{figure}[h!]
     \centering
         \centering
         \includegraphics[width=0.45\textwidth]{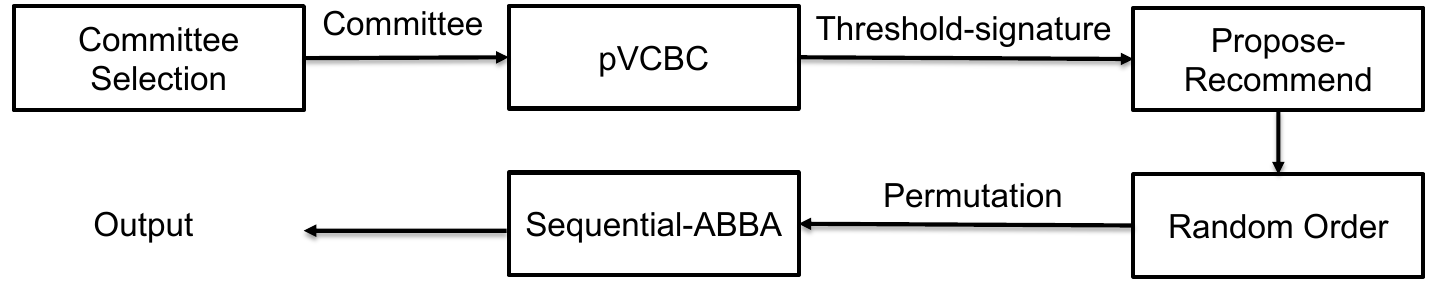}
         \caption{An overview of the proposed protocol framework.}
         \label{fig:Framework}
\end{figure}

\section{Design of cMVBA}

\subsection{Overview}
The cMVBA protocol achieves multi-valued Byzantine agreement by tolerating up to \( f < \frac{n}{3} \) Byzantine faults, with an expected communication complexity of \( O(ln^2 + \lambda n^2) \) and a constant asynchronous round count. The protocol is divided into five key sub-protocols: Committee Selection (CS), Prioritized Verifiable Consistent Broadcast (pVCBC), Propose-Recommend, Random Order, and Sequential-ABBA. Figure \ref{fig:Framework} illustrates the protocol's framework.

\subsection{Committee Selection (CS)}
CS optimizes Byzantine agreement efficiency by selecting a subset of \( f+1 \) parties, based on party ID and protocol instance, to broadcast proposals rather than requiring all \( n \) parties. This subset selection ensures that at least one honest party participates and is dynamically determined via a cryptographic coin-tossing scheme inspired by FasterDumbo \cite{FASTERDUMBO}. This randomness reduces risks of adversarial influence, prevents starvation, and mitigates Denial-of-Service attacks, ensuring fair participation across parties.

\subsection{Prioritized Verifiable Consistent Broadcast}
After committee selection, each member provides a verifiable proof of their proposal, confirming it has been sent to at least \( f+1 \) honest parties. This proof, a threshold signature, ensures consistency by verifying that the proposal has been disseminated accurately. A modified version of the traditional VCBC protocol, called pVCBC, is used here to limit broadcast only to committee members, improving efficiency.

\subsection{Propose-Recommend}
In cMVBA, the Propose-Recommend step replaces the classic commit step to enhance efficiency. Here, only the committee members propose requests, attaching threshold signatures as verifiable proofs. The recommend phase collects these proofs, reaching agreement once \( n-f \) recommendation messages are gathered. This streamlined process avoids the need for an \( n \)-length proof array, reducing complexity.

\subsection{Random Order}
This step generates a random permutation of the selected committee members to define proposal order, maintaining fairness and preventing adversaries from manipulating message delivery order. By randomizing order, cMVBA reduces the risk of increased asynchronous rounds and enhances protocol security.

\subsection{Sequential-ABBA}

The final phase in the cMVBA protocol is the Sequential-ABBA protocol, which is responsible for reaching an agreement on one of the proposals submitted by the selected parties. Building on the random order generated in the previous step, the Sequential-ABBA protocol runs an agreement loop that systematically evaluates each selected party's proposal. The input to this protocol consists of the permutation list generated in the Random Order step and the list of recommendations obtained from the Propose-Recommend step. The output is the final proposal agreed upon by the parties.

\section{Security Analysis of the Proposed Protocol}
\label{sec:eval}







\label{subsection:security}

The correctness of the proposed protocol is critical to ensure that it adheres to the standard MVBA outcomes. Our goal is to prove that reducing the number of broadcasts does not compromise the protocol's reliability. The output of a party's request depends on the output of an ABBA instance within our protocol's agreement loop, typically resulting in a value of \textit{1}. For the ABBA protocol to output \textit{1}, at least one honest party must input \textit{1}. We substantiate our protocol's integrity by demonstrating that at least $2f+1$ parties receive a verifiable proof of a party's proposal, thus ensuring the necessary input for the ABBA instance. Here we provide the lemmas and theorem to prove the effectiveness and correctness of the protocol inspired from \cite{PMVBA, OHBBFT, SlimABC}.

\begin{lemma}
\label{lemma1}
At least one party's proposal reaches $2f+1$ parties.
\end{lemma}

Since the selected parties can be byzantine and the adversary can schedule the message delivery to delay the agreement, we have considered the following scenarios:
\begin{enumerate}
    \item Among $\langle f+1 \rangle$ selected parties, $f$  parties are non-responsive.
    \item Selected $\langle f+1 \rangle$ parties are responsive, but other $f$ non-selected parties are non-responsive.
    \item Every party is responsive, including the selected $\langle f+1 \rangle$ parties.
    \item Selected $t \leq \langle f+1 \rangle$ parties are responsive, and total $m$ parties are responsive, where $\langle 2f+1 \rangle \leq m \leq n $.
    
\end{enumerate}

This lemma ensures that if $2f+1$ parties receive a verifiable proof, at least one honest party will input \textit{1} to the ABBA instance, ensuring the protocol reaches an agreement on \textit{1}.

\begin{lemma}
\label{lemma2}
Without any permutation, the adversary can cause at most $f+1$ iterations of the agreement loop in the Sequential-ABBA protocol.
\end{lemma}
 \begin{lemma}
 \label{lemma3}
 Let $\overline{A} \subseteq \{1, 2, ..., f+1\}$ be the set of selected parties for which at least $f+1$ honest parties receive the verifiable proof, and let $\Pi$ be a random permutation of the $f+1$ selected parties. Then, except with negligible probability:
 \begin{itemize}
     \item For every party $p \in \overline{A}$, the ABBA protocol on $ID|p$ will decide $1$.
     \item $|\overline{A}| \geq 1$.
     \item There exists a constant $\beta > 1$ such that for all $t \geq 1$, $Pr[\Pi[1] \notin \overline{A} \wedge \Pi[2] \notin \overline{A} \wedge ... \wedge \Pi[t] \notin \overline{A}] \leq \beta^{-t}$.
 \end{itemize}
\end{lemma}

 \begin{theorem}
 Given a protocol for biased binary Byzantine agreement and a protocol for verifiable consistent broadcast, the Prioritized-MVBA protocol provides multi-valued validated Byzantine agreement for $n > 3f$ and invokes a constant expected number of binary Byzantine agreement protocols.
 \end{theorem}





\section{Conclusion}
This paper introduced cMVBA, a novel MVBA protocol that minimizes communication complexity via a committee-based approach and ABBA integration. By dynamically selecting \(f+1\) parties to broadcast, cMVBA maintains resilience and efficiency, achieving optimal message complexity of \(O(n^2)\) and expected constant asynchronous rounds. The introduction of the pVCBC protocol to reduce the number of broadcasts further enhances the effectiveness of the protocol, making cMVBA ideal for large-scale decentralized applications. Future work will explore adaptive scalability, protocol optimizations, and empirical validations to enhance robustness in diverse settings.



\bibliography{references}



\end{document}